\documentclass[aps,prb,superscriptaddress,showpacs,twocolumn,letterpaper]{revtex4}

\usepackage{graphicx}
\usepackage{bm}

\begin{document}


\title {Luttinger liquid ARPES spectra from samples of
  Li$_{0.9}$Mo$_6$O$_{17}$ grown by the temperature gradient flux
  technique}

\author {G.-H. Gweon$^{\dagger}$}
\author {S.-K. Mo}
\author {J. W. Allen}

\affiliation {Randall Laboratory of Physics, University of Michigan, 500
  E. University, Ann Arbor, MI 48109}

\author {J. He}

\affiliation {Department of Physics and Astronomy, University of
  Tennessee, Knoxville, TN 37996}

\author {R. Jin}

\affiliation {Solid State Division, Oak Ridge National Laboratory, Oak
  Ridge, TN 37831}

\author {D. Mandrus} 

\affiliation {Department of Physics and Astronomy, University of
  Tennessee, Knoxville, TN 37996} 
\affiliation {Solid State Division, Oak
  Ridge National Laboratory, Oak Ridge, TN 37831}

\author {H. H\"ochst} 

\affiliation {Synchrotron Radiation Center, University of Wisconsin,
  Stoughton, WI 53589}

\date {\today}

\begin {abstract}
  
  Angle resolved photoemission spectroscopy line shapes measured for
  quasi-one-dimensional Li$_{0.9}$Mo$_6$O$_{17}$ samples grown by a
  temperature gradient flux technique are found to show Luttinger liquid
  behavior, consistent with all previous data by us and other workers
  obtained from samples grown by the electrolyte reduction technique.
  This result eliminates the sample growth method as a possible origin of
  considerable differences in photoemission data reported in previous
  studies of Li$_{0.9}$Mo$_6$O$_{17}$.

\end {abstract}

\pacs {71.10.Pm, 71.20.-b, 79.60.-i}
\maketitle

Li$_{0.9}$Mo$_6$O$_{17}$, also known as the Li purple bronze, is a
quasi-one-dimensional metal which displays metallic T-linear resistivity
and temperature independent magnetic susceptibility for temperatures down
to $T_X \approx 24$~K, where a phase transition of unknown origin is
signaled by a very weak anomaly in the specific heat \cite{Schlenker1985}.
As $T$ decreases below $T_X$, the resistivity increases.  However the
d.c.\ magnetic susceptibility is unchanged below $T_X$
\cite{Schlenker1985,matsuda86}, implying no single particle gap opening, 
and infrared optical studies \cite{degiorgi-purple-optics} below $T_X$
also show no gap opening down to $1$~meV. Consistent with this evidence
for the lack of a single particle gap, repeated x-ray diffraction studies
\cite{pouget-Li-struct} show no charge density wave or spin density wave.

The various transport and spectroscopy studies of this fascinating
material have been made on samples prepared by two methods, an electrolyte
reduction technique \cite{Schlenker1985} and a temperature gradient flux
technique \cite{greenblatt84}.  Angle resolved photoemission spectroscopy
(ARPES) is the only measurement for which any major inconsistency in data
obtained from samples prepared by the two different methods has been
reported, and the inconsistency is very serious.  In particular an
extensive set of ARPES data from two groups
\cite{grioni,denlinger,gweon1,gweon2,gweon3,allen,gweon4} obtained on
electrolyte reduction samples show non-Fermi liquid ARPES line shapes
consistent\cite{denlinger,gweon2,gweon3,allen,gweon4} with Luttinger
liquid (LL) behavior and no low temperature Fermi energy ($E_F$) gap,
whereas ARPES data reported \cite{xue,smith_reply,smith} for
temperature-gradient-flux 
grown samples show Fermi liquid (FL) line shapes, a large low temperature
$E_F$ gap and an additional feature inconsistent with the known band
structure of the material.  These differences between the two ARPES data
sets are summarized in Ref.\ [\onlinecite{gweon1}].  The LL line shapes
have been verified repeatedly in subsequent studies
\cite{gweon1,gweon2,gweon3,allen,gweon4} of samples prepared with the
electrolyte reduction technique.  Nonetheless it has been a lingering
possibility that FL line shapes and a large low temperature gap could
perhaps be characteristic of temperature gradient flux grown samples.
This Brief Report dispels that possibility by reporting ARPES spectra for
temperature gradient flux grown samples that are in full agreement with
the line shapes obtained for electrolyte reduction samples.

The spectra reported here were obtained on the PGM beamline at the
Wisconsin Synchrotron Radiation Center.  Photons of energy 30 eV were used
to excite photoelectrons whose kinetic energies and angles were analyzed
with a Scienta SES 200 analyzer.  Measurement on a freshly prepared Au
surface was used to determine the position of $E_F$ in the spectra and the
overall energy resolution of 21 meV due to both the monochromator and the
analyzer.
The angle resolution was set at $\pm 0.1^\mathrm{o}$, better
than that $\pm 0.25^\mathrm{o}$ in our earlier work \cite{gweon1} and
exactly the same as used in previous ARPES studies
\cite{xue,smith_reply,smith} of temperature gradient flux samples.
The sample surface was obtained by cleaving {\em in situ}
and the data were taken at a sample temperature of 200 K, much higher than
the transition temperature 24 K\@.

For the endstation in place at the time of taking the data reported here,
the angular dispersion direction of the SES 200 analyzer was vertical.
ARPES symmetry analysis of the data obtained shows that the one
dimensional $\Gamma$--Y chain axis direction was (unintentionally)
oriented at an angle of 13$^\mathrm{o}$ to the vertical.  Nonetheless we
will refer to this geometry as the ``vertical geometry'' from here on in
the paper.  Due to this small angular offset, the dispersions in this data
set are slightly different from those that we obtained previously along
the $\Gamma$--Y axis, as documented carefully in discussing Fig.\ 2 below.
We have repeated the measurement in exactly the same geometry as that of
our previous experiments
\cite{denlinger,gweon1,gweon2,gweon3,allen,gweon4}, i.e.\ one in which
both the one dimensional chain axis and the angular dispersion direction
of the analyzer are horizontal and well aligned, and found dispersions
essentially identical to those of the previous
\cite{denlinger,gweon1,gweon2,gweon3,allen,gweon4} data.  We will refer to
this geometry as the ``horizontal geometry'' below.  Despite the small
angular offset, we present here the data taken in the vertical geometry
because (1) these data happen to show the $E_F$-crossing line shapes most
clearly among all of our data sets, by virtue of having fortuitously the
maximum intensity of the band crossing $E_F$ relative to the intensities
of the bands that do not cross $E_F$, and (2) the differences in
dispersions have been verified to arise from the small offset and are in
any case so slight as to be insignificant for the central thrust of this
paper.  Both of these points are elaborated below.
Another advantage of the vertical geometry setup was that it allowed
acquisition of intensity maps like the one presented below (Fig. 1) for
many parallel one dimensional paths crossing the FS.  Thereby we could
verify that the LL behavior holds for such paths anywhere in the Brillouin
zone, regardless of the exact location of the momentum space cut across
the Fermi surface, so that Fermi liquid behavior does not occur for some
very specific cut, as reported previously \cite{xue,smith_reply} for
temperature gradient flux samples.

\begin{figure}
\includegraphics{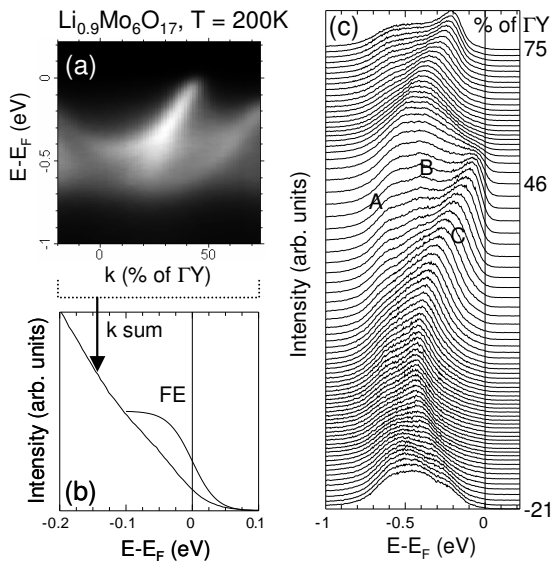}
\caption{ARPES data obtained for samples of Li$_{0.9}$Mo$_6$O$_{17}$ prepared by
  the temperature gradient flux growth.  (a) k-energy map where k is the
  momentum projected onto the $\Gamma$--Y axis.  (b) The k-sum of the data
  in (a).  A Fermi edge (FE) spectrum for the employed energy resolution
  (21 meV FWHM) and the measurement temperature (200K) is shown to
  demonstrate the non-FE nature of the k-sum.  (c) An energy distribution
  curve (EDC) stack representation of the data shown in (a).  The spectrum
  corresponding to k$_F$ is drawn with a thick line.  The momentum
  increment is 1.3\% of $\Gamma$--Y.}
\end{figure}

Fig.\ 1 shows ARPES spectra taken in the vertical geometry on a sample
grown by the temperature gradient method.  Panels (a) and (c) summarize
the overall electronic structure with k labels denoting the k values
projected onto the $\Gamma$-Y axis.  We label the bands A,B,C in the
order of decreasing binding energy at $\Gamma$.  As we will discuss in
connection with Fig.\ 2 below, the overall band structure revealed by the
data is consistent with the data in the literature
\cite{grioni,denlinger,gweon1,gweon2,gweon3,allen,gweon4} obtained on
samples grown by the electrolyte reduction method, as well as with band
theory \cite{whangbo}.  Furthermore, the LL line shapes observed
previously \cite{denlinger,gweon1,gweon2,gweon3,allen,gweon4} are not just
confirmed, but actually better observed due to the enhanced strength of
band C relative to that of bands A,B in the present data.  For example, we
can now clearly observe that the spectral weight of band C shows a
back-bending behavior after the peak has crossed the Fermi level (darker
curve), one of the key signatures of the LL line shape. In panel (b) we
show the k-sum of the ARPES data.  As found previously, the resulting line
shape is far from the Fermi edge line shape expected of a FL and instead
is described much better by a LL with $\alpha > 0.5$, where $\alpha$ is
the so-called the anomalous dimension of the LL.

Panel (a) of Fig.\ 2 summarizes the overall band structure determined from
the present data (open circles) and compares it with that from our previous
result (diamonds) taken on a sample grown by the electrolyte reduction
technique.  The small differences arising from the slightly different
k-paths can be seen.  For example, in the new data band B becomes almost
non-dispersive when peak C crosses $E_F$ while this occurs for larger k
values for the data perfectly along $\Gamma$-Y.  As shown in panel (b)
band theory predicts bands A, B, and C essentially as observed, and also a
fourth band D.  Bands A and B do not cross $E_F$ and C and D become
degenerate and cross $E_F$ together. All four bands have been observed
\cite{denlinger,gweon1,gweon2} for various k-paths, although band D is
typically very weak, just a slight shoulder on the leading edge of peak C,
and is clearly seen only for a particular k-path \cite{denlinger,gweon3}
where it appears as a main peak.  In the vertical geometry data, band D is
nearly undetectable (see Fig.\ 1 (c)) but was observed very weakly in the
horizontal geometry data, consistent with previous results.  For
completeness, we mark the approximate position of band D thus found for
the present sample as a gray region.  This position is similar to that
found for previous samples along the same, i.e.\ the $\Gamma$-Y, direction
as well as along the special k-path \cite{denlinger,gweon3} where D is
strong.

\begin{figure}
\includegraphics{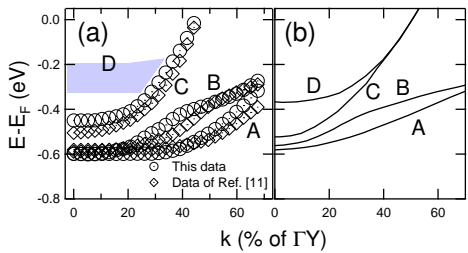}
\caption{Agreement of the overall band structures obtained on the two
  samples grown with different methods.  (a) Momentum-energy dispersion
  relations as extracted by taking the peak positions of energy
  distribution curves (e.g.\ in Fig.\ 1 (c)).  The data plotted in circles
  correspond to the data of the current sample (Fig.\ 1) grown by the
  temperature gradient flux method and the data plotted in diamonds
  correspond to the data reported in Ref.\ [\onlinecite{gweon3}] for
  sample grown by the electrolyte reduction method.  See text for
  discussion of small differences visible, arising from slightly different
  k-paths.  The approximate position of the band D for both samples is
  indicated as a gray region.  (b) Extended H\"uckel tight binding band
  structure calculation \cite{whangbo} for comparison.  Note that the
  energy scale of the calculation was multiplied by a factor of 2.2 in
  order to roughly match the dispersion of the experiment.}
\end{figure}

Fig.\ 3 compares the $E_F$ crossing line shapes measured on the samples
grown by the two different methods, with panel (a) showing the new data in
the vertical geometry for the temperature gradient flux sample, and panel
(b) showing data from Ref.\ [\onlinecite{gweon3}] taken at the same photon
energy for the electrolyte reduction sample.  In each panel, the data are
presented with the spectra for the various k-values overplotted to better
show the approach and $E_F$ crossing of peak C.  As explained in the
previous paragraph, a small difference of the band B dispersion arises
from the slightly different k-paths.  The general features of the two sets
of spectra are nearly identical, except that, as mentioned already, in (a)
the strength of band C relative to that of the non-$E_F$ crossing band B
is greater than in (b).  Therefore the intrinsic line shape features of
band C, which we have shown \cite{denlinger,gweon2,gweon3,allen,gweon4} to
be well described by the LL line shape theory, are now even more clearly
visible.  These include the spinon edge and the holon peak, which disperse
with different velocities, the diminution of intensity as $E_F$ is
approached, and the back dispersing edge after the peak has crossed $E_F$.
One is now forced to conclude that the large disagreement of the overall
band dispersions and $E_F$ crossing line shapes found previously
\cite{gweon1} for the ARPES data reported by Xue et al.\cite{xue,smith}
and those reported by ourselves and others
\cite{grioni,denlinger,gweon1,gweon2,gweon3,allen,gweon4} do not stem from
the sample growth method.

\begin{figure}
\includegraphics{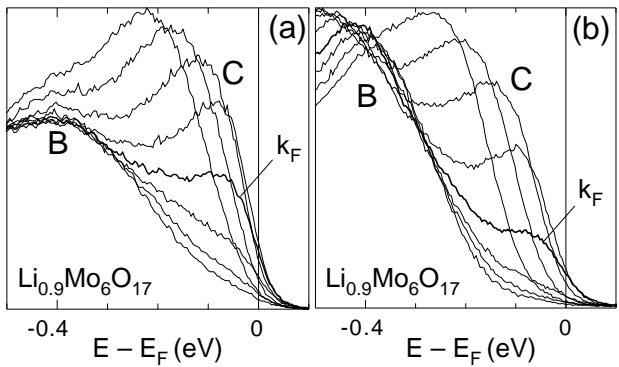}
\caption{Identical nature of LL ARPES lineshapes obtained for samples 
  of Li$_{0.9}$Mo$_6$O$_{17}$ prepared by (a) the temperature gradient
  flux growth and (b) electrolyte reduction methods.  The data in (b) is
  from Ref.\ [\onlinecite{gweon3}].  In both panels, the momentum
  increment is 2.6\% of $\Gamma$--Y.}
\end{figure}

Before concluding, we note that samples prepared by us [JH, RJ and DM] in
the same way as for those used in the ARPES reported here, have also been
used for new measurements of the temperature dependences of the
resistivity, specific heat, magnetic susceptibility and optical properties
\cite{musfeldt}.  These results have re-confirmed that no gap opening is
associated with the low $T$ resistivity rise and have been interpreted as
showing the probable importance of localization effects for the properties
below $T_X$.  Although the lower energy limit of the new optical study is
10 meV, larger than the minimum energy of 1 meV of a previous optical
study \cite{degiorgi-purple-optics}, it is nonetheless smaller than the
energy resolutions used in any ARPES studies on the material to date
($\geq 15$ meV).  Further, the new optical study found that the spectral
weight {\em increases} below $T_X$ in the low energy sector ($< 100$ meV)
for which previous ARPES studies \cite{xue,smith_reply,smith} on
temperature gradient flux samples found a large gap opening (2$\Delta \approx$
80 meV).

To summarize, we have shown that the ARPES spectra of
Li$_{0.9}$Mo$_6$O$_{17}$ samples prepared by temperature gradient flux
growth display LL behavior the same as seen for samples prepared by the
electrolyte reduction method, thus augmenting further the strong case for
LL ARPES lineshapes already established by our past ARPES work on this
material.

\begin{acknowledgments}
  
  This work was supported by the U.S. NSF grant DMR-99-71611 and the U.S.
  DoE contract DE-FG02-90ER45416 at U. Mich.  The ORNL is managed by
  UT-Battelle, LLC, for the U.S. DoE under contract DE-AC05-00OR22725.
  Work at UT was supported by the NSF Grant DMR 00-72998.  The SRC is
  supported by the NSF Grant DMR-0084402.

\end{acknowledgments}

$^{\dagger}$ Current address: MS 2-200, Lawrence Berkeley National
Laboratory, 1 Cyclotron Road, Berkeley, CA, 94720; Electronic address:
gweon@umich.edu.

\end{document}